\documentclass[onecollarge,natbib]{svjour2}
\bibpunct{[}{]}{;}{n}{}{,} 
\smartqed  
\usepackage{graphicx}
\journalname{Few-Body Systems}

\usepackage[utf8]{inputenc}
\usepackage{amssymb}
\usepackage{graphicx}
\usepackage{float}
\usepackage[colorlinks,linkcolor=blue]{hyperref}
\usepackage{caption}
\usepackage{color}

\newcommand{\beq}{\begin{eqnarray}}
\newcommand{\eeq}{\end{eqnarray}}
\newcommand{\nneeq}{\nonumber \end{eqnarray}}

\newcommand{\es}{& = &}

\newcommand{\rs}{\, = \,}

\newcommand{\cM}{ {\cal M} }
\newcommand{\cH}{ {\cal H} }

\newcommand{\cG}{ {\cal G} }
\newcommand{\cT}{ {\cal T} }

\newcommand{\cL}{ {\cal L} }

\begin{document}

\title{ Effective particles in quantum field theory }

\author{Stanis{\l}aw D. G{\l}azek \and Arkadiusz P. Trawi\'nski }
\institute{S. D. G{\l}azek \at Department of
Physics, Yale University, 217 Prospect Street, New Haven, CT 06511-8499; 
on leave of absence from Faculty of Physics, University of Warsaw, Pasteura 5, 
02-093 Warszawa, Poland\\
\email{Stanislaw.Glazek@yale.edu}
\and
A. P. Trawi\'nski \at 
Faculty of Physics, University of Warsaw, Pasteura 5, 
02-093 Warszawa, Poland\\
\email{Arkadiusz.Trawinski@fuw.edu.pl}}

\date{ 29 November 2016 }

\maketitle

\begin{abstract}
The concept of effective particles is introduced 
in the Minkowski space-time Hamiltonians in quantum 
field theory using a new kind of the relativistic 
renormalization group procedure that does not 
integrate out high-energy {\it modes} but instead 
integrates out the large {\it changes} of invariant
mass. The new procedure is explained using examples 
of known interactions. Some applications in 
phenomenology, including processes measurable in 
colliders, are briefly presented.

\end{abstract}

\section{ Introduction } 

\begin{figure}[ht]
\centering
\includegraphics[width=.8\textwidth]{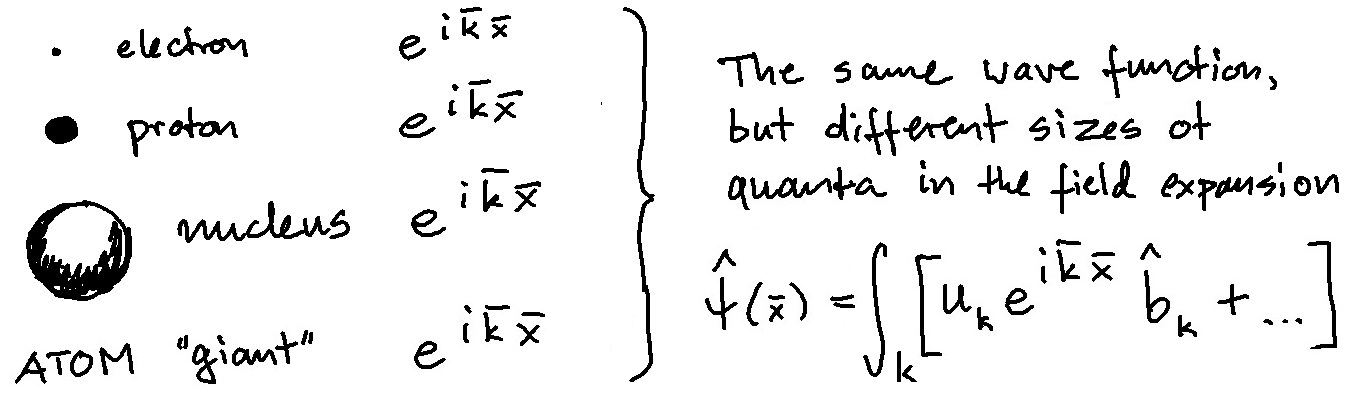}
\caption{Quantum field operators are blind to the size of individual quanta.}
\label{fig:size}
\end{figure}

Effective particles can be introduced in the quantum 
field theory (QFT) in such a way that their size plays 
the role of a scale parameter in a renormalization 
group procedure for the corresponding Hamiltonians. 
The example of QCD is particularly interesting because 
the effective quarks of large size are expected to 
correspond to the constituent quarks that are used in 
classification of hadrons, while point-like quarks and 
gluons correspond to partons used in description of 
hadrons in high-energy collisions. But the renormalization 
group procedure for effective particles (RGPEP) that is 
used here is not limited to QCD. Its general nature stems 
form the fact that one does not specify the size of quanta 
when one expands a quantum field in space into its Fourier 
modes. Namely, the Heisenberg and Pauli~\cite{Ref1,Ref2} 
formulation of QFT is equally applicable to various 
particles despite the fact that they have different 
sizes, see Fig.~\ref{fig:size}. One can say that the 
concept of a quantum field operator is blind to the 
size of individual quanta. In contrast, in the renormalized 
theories derived using the RGPEP, this free size parameter 
determines the width of vertex form factors in the 
Hamiltonian interaction terms for effective particles.  

\section{ Example of the RGPEP calculation: asymptotic freedom in Yang-Mills theory }
\label{sec:YM}

\begin{figure}[ht]
\centering
\includegraphics[width=0.8\textwidth]{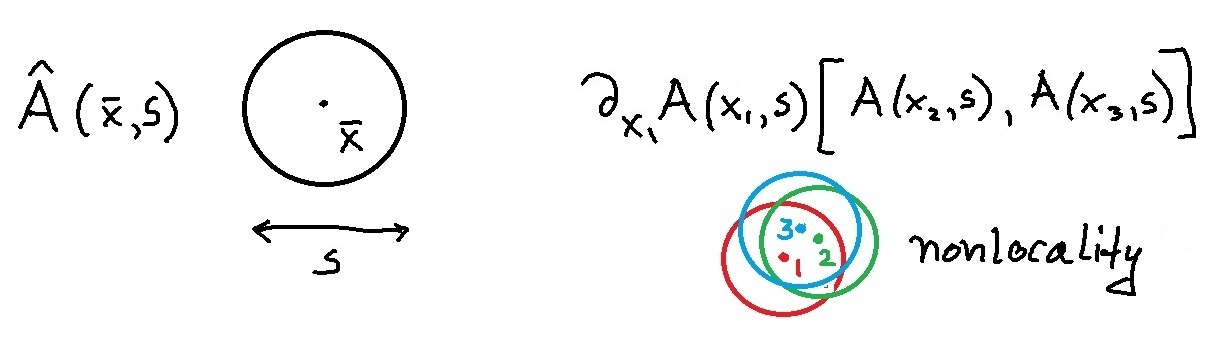}
\caption{ Nonlocal interaction of quanta of size $s$ 
can be approximated by a local interaction for wavelengths 
much greater than $s$. The effective particle size $s$ is 
the free scale parameter in the RGPEP.}
\label{fig:Nonlocality}
\end{figure}

Canonical quantization of the Yang-Mills (YM) 
theory begins with the Lagrangian density 
${\cal L} = - \ {\rm tr} \, F^{\mu \nu}
F_{\mu \nu}/2 $, in which the field strength
tensor is $F^{\mu \nu} = \partial^\mu A^\nu - 
\partial^\nu A^\mu + i g [A^\mu, A^\nu]$.
The associated energy-momentum tensor is
$\cT^{\mu \nu} = -F^{a \mu \alpha} \partial^\nu 
A^a_\alpha + g^{\mu \nu} F^{a \alpha \beta} 
F^a_{\alpha \beta}/4$. Consequently,
the field four-momentum is given by
$P^\nu = \int_\Sigma d\sigma_\mu \ \cT^{\mu\nu}$,
where $d\sigma_\mu$ is a measure on the 
hypersurface $\Sigma$ in the Minkowski 
space-time. In the front form (FF) of Hamiltonian 
dynamics~\cite{Dirac}, one uses the variables $x^\pm 
= t \pm z$ and $x^\perp = (x,y)$, with a
similar convention for all tensors.
The FF Hamiltonian results from integration
of $\cH_{YM} = T^{+ \, -}$ over the front
hypersurface $\Sigma$ defined by the condition $x^+=0$.
The canonical FF Hamiltonian density in gauge 
$A^+=0$ contains three terms, $\cH_{YM} = 
{\cal H}_{A^2} + {\cal H}_{A^3} + {\cal H}_{A^4}$. 
Our example concerns the three-gluon term 
$\cH_{A^3}$, to exhibit asymptotic freedom.
Quantization is achieved by replacing the 
classical gauge-field $A$ by a field operator,
\beq
\label{quantization}
\quad 
\hat A^\mu \es \sum_{k \sigma c}
\left[ t^c \, \varepsilon^\mu_{k\sigma} \,
a_{k\sigma c} \, e^{-ikx} + t^c \,
\varepsilon^{\mu *}_{k\sigma} \,
a^\dagger_{k\sigma c} \, e^{ikx}\right]_{on ~
\Sigma } \ , 
\eeq
where $a_{k\sigma c}$ and $a^\dagger_{k\sigma c}$ 
are the annihilation and creation operators
of gluons with momentum $k=(k^+, k^\perp)$,
polarization vectors $\varepsilon_\sigma$ and 
color matrices $t^c$ that span the algebra of 
$SU(N_c)$. We consider the number of colors 
$N_c=3$. Thus, the starting Hamiltonian is 
\beq
\hat H_{YM} \es  \int_\Sigma  \
: \cH_{YM}(\hat A ) : \ ,
\eeq 
where the double dots indicate normal ordering.
Our focus is on the term that comes from 
${\cal H}_{A^3} =  g \ i\partial_\alpha A_\beta^a
\ [A^\alpha,A^\beta]^a$, which in the local quantum 
theory for point-like particles leads to 
\beq
\hat H_{A^3} \es  \int_{x \, \in \, \Sigma }  \
: g \ i\partial_\alpha \hat A_\beta^a(x)
\ [\hat A^\alpha(x), \hat A^\beta(x)]^a : \ .
\eeq 
If the quanta are not point-like, see
Fig.~\ref{fig:Nonlocality},
the local interaction is merely an approximation,
in which the size $s$ is a free parameter. Thus,
instead of the quantum size-blind field operator 
$\hat A(x)$, we can introduce an effective particle 
quantum field $\hat A(x,s)$ and the non-local
three-gluon interaction term
\beq
\label{gsx}
\hat H_{A^3, s} \es  \int_{x_i \, \in \, \Sigma }  \
g_s(x_1,x_2,x_3) \ : i\partial_\alpha A_\beta^a(x_1,s)
\ [A^\alpha(x_2,s),A^\beta(x_3,s)]^a : \ .
\eeq
The function $g_s(x_1, x_2, x_3)$ is determined 
using the RGPEP, except that one works with momentum 
variables rather than the space-time coordinates. 
The required elements of the RGPEP can be found 
in a sequence of Refs.~\cite{GlazekWilson1,Wilsonetal,
Wegner,RGPEP,GomezRochaGlazek}. A brief summary of 
the procedure follows.

The RGPEP employs equations that connect theories 
with different values of the effective particle 
size $s$. There is an initial condition set at 
$s=0$ by canonical QCD. In our example, 
\beq
\label{hatP-}
\hat H_{A^3, s=0} \es  \int_{x \, \in \, \Sigma }  \
: g \ i\partial_\alpha \hat A_\beta^a(x,0)
\ [\hat A^\alpha(x,0), \hat A^\beta(x,0)]^a : \ ,
\eeq
where the field operator is built from creation and annihilation  
operators for point-like quanta,
\beq
\label{A0}
\hat A^\mu(x,s=0) \es \sum_{k \sigma c} 
\left[ t^c \, \varepsilon^\mu_{k\sigma} \,
a_{k\sigma c, s=0} \, e^{-ikx} + t^c \,
\varepsilon^{\mu *}_{k\sigma} \,
a^\dagger_{k\sigma c, s=0} \, e^{ikx}\right]_{on ~
\Sigma } \ .
\eeq 
Direct evaluation yields
\beq
\label{H0}
\hat 
H_{A^3, s=0} \es
\sum_{123}
\ \delta_{12.3}  \  
\left[g\,Y_{123}\ a^\dagger_{1,0} \, a^\dagger_{2,0} \, a_{3,0} +
g\,Y_{123}^*\ a^\dagger_{3,0} \, a_{2,0} \,
a_{1,0} \right] \ ,
\eeq
where an abbreviated notation is used, such as 1 in place 
of $k_1, \sigma_1, c_1$, the factor $\delta_{12.3}$ takes 
care of momentum conservation and the coefficients $Y_{123}$ 
stand for the gluon momentum, spin and color dependent 
factors implied by Eqs.~(\ref{hatP-}) and (\ref{A0}).
This is an example of a term in the general initial 
Hamiltonian for $s=0$. The general structure of
the latter reads
\beq
H_0(a_0) \es
\sum_{n=2}^\infty \, 
\sum_{i_1, i_2, ..., i_n} \, c_0(i_1,...,i_n) \, \, a^\dagger_{i_1,0}
\cdot \cdot \cdot a_{i_n,0} \ ,
\eeq
where the coefficients $c_0$ are determined by the regulated
canonical theory and counterterms. In the same notation, the 
Hamiltonian for effective particles of size $s$ reads
\beq
\label{Hcs}
H_s(a_s) \es
\sum_{n=2}^\infty \, 
\sum_{i_1, i_2, ..., i_n}
  \, c_s(i_1,...,i_n) \, \, a^\dagger_{i_1,s}
\cdot \cdot \cdot a_{i_n,s} \ ,
\eeq
and the coefficients $c_s$ are calculated using
the RGPEP. One of these coefficients is $g_s(i_1, i_2, i_3)$
that corresponds to the function $g_s(x_1, x_2,
x_3)$ in Eq.~(\ref{gsx}).

The coefficients $c_s$ in Eq.~(\ref{Hcs})
are derived from the condition that the
Hamiltonian is not changed, $H_s(a_s) = H_0(a_0)$,
when the canonical quanta with size $s=0$ are replaced
by the effective quanta of size $s > 0$, using 
a unitary transformation, $a_s = U_s \, a_0 \, U_s^\dagger$.
One works in a canonical operator basis and
calculates the same coefficients $c_s$ using
the operator $\cH_s = H_s(a_0) = U_s^\dagger H_s(a_s) U_s$
that is determined by the RGPEP equations 
\beq
\label{RGPEP1}
\cH'_s \es \left[ \cG_s , \cH_s \right] \ , \\ 
\label{RGPEP2}
\cG_s  \es [ \cH_f, \tilde \cH_s ] \ ,
\eeq
where $\tilde \cH_s$ differs from $\cH_s$ only 
by multiplication of coefficients $c_s$ by the 
square of $+$-momentum carried by the corresponding
product of creation or annihilation operators.
This multiplication secures boost invariance of 
the effective theory. In Eqs.~(\ref{RGPEP1}) and
(\ref{RGPEP2}), the Hamiltonian is divided into 
a free part, equal to $\cH_0$ for $g=0$, and the 
interaction $\cH_I$, so that $\cH_s = \cH_f + 
\cH_I$. The key feature of solutions to these 
equations, and the reason for using them here, 
is that the coefficients $c_s$ exponentially suppress 
interactions that change the invariant mass of 
effective particles by more than $1/s$, see below. 
No modes are eliminated from the dynamics.
Instead, vertex form factors suppress changes
of invariant mass, see below.

Solutions to Eq.~(\ref{RGPEP1}) define the YM 
theories  {\it a priori} non-perturbatively. As
such, they are not easy to obtain. In contrast, 
asymptotic freedom is found in the expansion 
in powers of the coupling constant,
\beq
\cH_s     \es \cH_f + g \cH_{1s} + g^2 \cH_{2s} +
g^3 \cH_{3s} + ... \ .
\eeq 
The terms of order $g$ and $g^3$ yield
the result~\cite{GomezRochaGlazek}
\beq
H_{A^3 \, (1+3)s} \es 
\sum_{123}\int_{123} \ \delta_{12.3} \  
e^{-s^4 \, \cM_{12}^4 } \
\left[\,V_{s123}\, a^\dagger_{1,s} a^\dagger_{2,s} a_{3,s} 
      + V_{s123}^*\, a^\dagger_{3,s} a_{2,s}
a_{1,s} \right] \ ,
\eeq
in which $\cM_{12}$ denotes the invariant mass of
gluons 1 and 2. We see that the effective three-gluon
interaction vertex is softened by the exponential 
form factor, which suppresses the interaction that 
involves a pair of gluons with invariant mass $\cM_{12}$ 
the more the larger size $s$. The calculation shows 
that in the limit of vanishing relative transverse 
momentum of gluons $V_{s123} = g_s \ Y_{123}$ and  
\beq
g_s \es
g_0 + { g_0^3 \over 48 \pi^2 }   N_c \,   11 \,
\ln { s \over s_0} \ .
\eeq
The resulting Hamiltonian function $\beta(s)$ 
corresponds to the Green's function $\beta(\lambda)$
obtained in Refs.~\cite{GrossWilczek,Politzer},
where $\lambda$ denotes the length of a four-dimensional
Euclidean momentum. Thus the effective particle 
formulation of QFT shows how the key feature of 
four-dimensional calculations of QCD manifests 
itself in the renormalized Hamiltonians. This way, 
the effective Hamiltonian approach passes an
obligatory test for tackling issues of strong 
interaction theory. It provides the Minkowskian 
space-time scale parameter that corresponds to 
the abstract four-dimensional Euclidean scale 
parameter: the size of effective particles.  
 
\section{ Early examples of RGPEP insight in phenomenology and theory of particles }

The RGPEP provides a conceptual insight into
particle dynamics in various systems. For example,
in the case of protons, the universal RGPEP
parameter $s$ tells us that the low-energy
effective theory that can be sought using QCD, 
to support the quark model classification of hadrons, 
concerns quarks whose size is comparable with the 
size of the proton itself. Such large quarks ought 
to be thought about as built from partons of smaller 
size $s$, and the RGPEP provides mathematical tools
for the corresponding construction of operators and
states in the effective Fock-space basis. Using this 
insight, one can ask if the configurations of three bulky 
effective quarks built of small partons can explain 
ridge-like correlations in high-energy $pp$ collisions 
observed by the ATLAS and CMS collaborations at 
LHC~\cite{ATLAS,CMS}. It was found~\cite{Kubiczek} 
that among various simple models, including the gluon 
string stretched between a quark and a diquark~\cite{BBG}, 
three quarks joined by a three-prong gluon body in the 
shape of letter Y, or a combination of various shapes 
and distributions of partons in effective particles,
only the Y effective-particle picture of a proton
provides eccentricity increasing with
multiplicity. The model study~\cite{Kubiczek} was
carried out before the data exhibiting such type
of behavior of $\sqrt{s}$ = 13 TeV $pp$
collisions~\cite{ATLAS,CMS} were published. This
example illustrates the phenomenological utility 
of the scale dependent, effective particle picture
in QFT.

The RGPEP can be applied to QED in description of
hydrogen and muonic hydrogen atoms~\cite{radius}
for the purpose of explanation of the difference
between the proton radii extracted from data for
level splittings in these
atoms~\cite{radiusreview}. This is a
computationally ambitious goal. But the question
of principle that RGPEP answers is how to derive
the Schr\"odinger equation from relativistic
QFT. In QED, the RGPEP would
start from the classical Lagrangian density $ \cL
= - {1 \over 4} F_{\mu \nu} F^{\mu \nu} + \bar
\psi( i \partial \hspace{-5pt}/ - e A
\hspace{-5pt}/ - m) \psi$. One treats the electrons 
and protons as point-like and proceeds to
quantization as in Sec.~\ref{sec:YM} to obtain
$\hat H$, regularize it, identify counterterms and
calculate $\hat H_{QED \, s}$. The eigenvalue problem,
$ \hat H_{QED \, s} |\psi_s \rangle = E |\psi_s
\rangle $ is identified with the Schr\"odinger
equation for the hydrogen atoms represented as
lepton-proton bound states when the size parameter
$s$ is sufficiently large to prevent the interactions
from inducing the invariant mass changes that exceed 
the constituent masses. The RGPEP valence constituent 
picture of atoms is illustrated schematically by
\beq
\left[
\begin{array}{c}
.. \\
.. \\
e \gamma e \bar e p \gamma \\
e \gamma p \gamma \\
e \gamma p  \\
\fbox{ $ e p $ }
\end{array}
\right]
\es
\left[
\begin{array}{c}
.. \\
.. \\
e \gamma e \bar e \\
e \gamma \gamma \\
e \gamma \\
e 
\end{array}
\right]
\times 
\left[
\begin{array}{c}
.. \\
.. \\
p e \bar e \\
p \gamma \gamma \\
p \gamma   \\
p 
\end{array}
\right] + ... 
\rs
|e_s p_s \rangle + ... \ .
\eeq
The left column represents a hydrogen atom as a 
superposition of virtual particle components in 
canonical theory, the size of quanta $s=0$. The 
lowest Fock component is put in a frame to indicate 
the state that provides the first approximation 
to an atom when one works with canonical QED.
Once one introduces the effective quanta, the 
whole tower of Fock components is rewritten as
an effective state of $| e_s p_s \rangle$ plus
more effective-particle components, indicated by the 
three dots. When the size parameter $s$ is increased,
the canonical interactions that act in and among 
all the Fock components are increasingly transformed 
into a complex interaction that predominantly 
acts in the component $| e_s p_s \rangle$,
because the change of the number of quanta,
even massless photons, involves much larger
changes of invariant mass than the changes
that occur in the component $| e_s p_s \rangle$.
For sufficiently large $s$ and for small $\alpha$,
the eigenvalue equation $ \hat H_{QED \, s} 
|\psi_s \rangle = E |\psi_s \rangle $ can be 
reduced to the eigenvalue equation for the 
component $| e_s p_s \rangle$, which, after 
including proton electromagnetic form factors,
especially the electric one, $G_E$, takes the 
familiar Schr\"odinger form up to terms of
formal order $\alpha$, 
\beq                
\label{Hpsi=EpsiRGPEP}
{ \vec p\, ^2 \over 2\mu} \, \psi_s(\vec p \,)
+ \int {d^3 k \over (2\pi)^3 } \, 
V_s(\vec p, \vec k\,) \, \psi_s(\vec k\,)
\es 
-E_B \, \psi_s(\vec p\,) \ ,
\eeq
where the potential is 
\beq
\label{V}
V_s(\vec p, \vec k \, )
\es 
e^{ - s^4 ( \vec p^{\, 2} - \vec k^{\, 2})^2/c^4 } 
\ \
{- 4\pi \alpha \over \vec q\,^2 } 
\ \ G_E(\vec q \,^2)  \ .
\eeq
The three-momentum transfer is $\vec q = \vec p -
\vec k$. The exponential factor that corrects the
Coulomb potential of extended proton, suppresses
the interaction that changes the invariant mass 
of effective electron-proton system by more than 
$1/s$. Once the size $s$ is selected to secure 
universal scaling of atomic levels with $\alpha$,
setting $s \sim 1/\sqrt{\mu M}$, where $\mu$ is
the reduced and $M$ the average constituent mass,
this factor introduces a small dependence of the
potential on the lepton mass. This minuscule 
dependence has a divergent perturbative expansion 
in powers of $\alpha$ when one expresses momenta
in units of $\alpha \mu c$. Therefore, it needs to 
be evaluated numerically. It turns out that it is 
capable of producing a lepton-mass correction in 
the extraction of proton radius from atomic levels, 
of the same magnitude~\cite{radius} as the variation 
encountered in the proton radius puzzle~\cite{radiusreview} 
when the effective nature of particles appearing 
in the Schr\"odinger equation according the RGPEP 
is not accounted for. 

It is worth pointing out that the Schr\"odinger
equation emerges from the RGPEP derivation of 
effective low-energy Hamiltonian in QFT using 
the three-momentum variables $\vec p = (p^\perp, p^z)$,
\beq
\label{kperp}
p^\perp
\es 
c \
\left[ (1-x) \ p_l^\perp - x \ p^\perp_p \right]
/\sqrt{ x(1 - x) } \ , \\
\label{kz}
p^z
\es
c \
(m_l + m_p) \ (x - \beta) /\sqrt{ x(1 - x) } \ ,
\eeq
where $\beta = m_l/(m_p+m_l)$, $c = \sqrt{m_l m_p} 
/(m_l + m_p)$, $x=p^+_l/(p^+_l + p^+_p)$, while $l$ 
refers to lepton and $p$ to proton. These
variables correspond to the relative momentum
variables used in light-front
holography~\cite{BrodskyPhysRep} for description of
hadrons. They appear to be non-relativistic while
they are in fact the relative momenta of
constituents in a relativistic theory, in which
boost invariance is maintained. The correspondence
is unlikely to be a meaningless coincidence. The
Schr\"odinger equation is universally valid as an
effective theory for atomic physics, despite that
it apparently contains no remnants of complexity
of relativistic QED. Similarly, the light-front
holography~\cite{BrodskyPhysRep}, motivated by the
idea of duality~\cite{Maldacena}, is meant to
approximate complex QFT in terms of solutions to
relatively simple field equations in AdS~\cite{adsqft}. 
In the case of strong interactions, the same variables 
are also used in showing that the linear confining 
potential in the instant form of dynamics~\cite{Dirac} 
corresponds to a quadratic one in the FF~\cite{Trawinski}. 

\section{ Jet production in pion-nucleus collisions }
\label{jets}

\begin{figure}[ht]
\centering
\includegraphics[width=.6\textwidth]{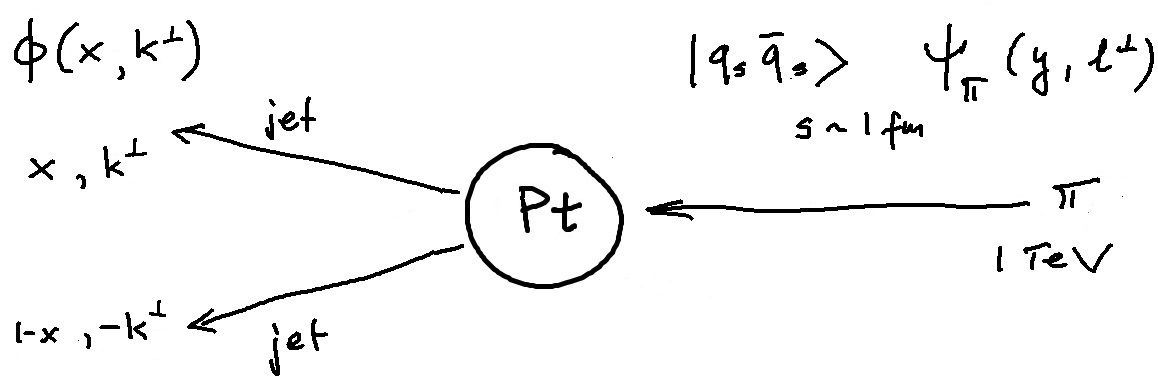}
\caption{ Pion is split into two quark jets on the gluons in a nucleus
          that act like a wedge. }
\label{fig:pionA}
\end{figure}

When a pion with energy on the order of TeV
collides with a platinum target, as in the E791 
Fermilab experiment~\cite{pionjets}, and
diffractively dissociates into two jets, the
observed jet distribution is expected to report 
on the quark-antiquark pion wave function.
The collision is illustrated in Fig~\ref{fig:pionA}.
Our description of it is based on Ref.~\cite{thesis}.

\begin{figure}[ht] 
\centering
\includegraphics[width=.5\textwidth]{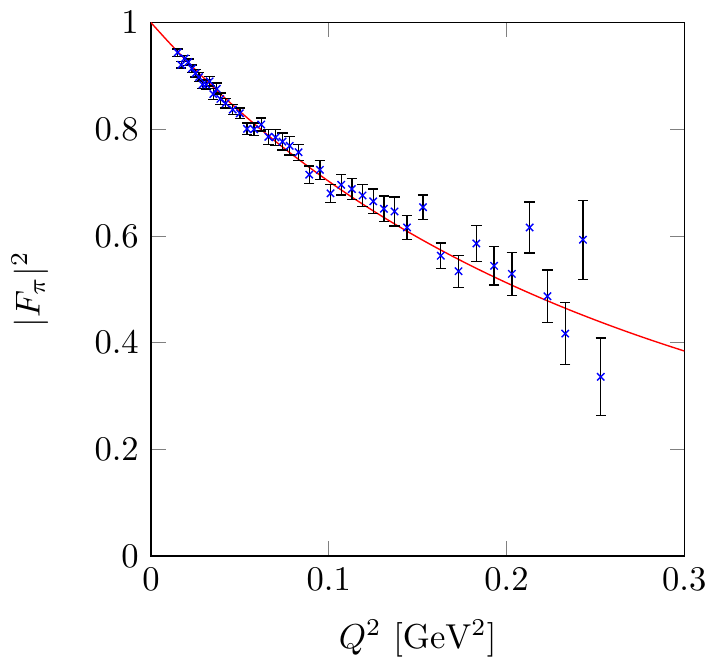} 
\caption{ The pion electromagnetic form factor at
low momentum transfers for the wave function model 
with holographic effective potential $U_{\rm eff} 
= \varkappa^4 r^2/4$ in AdS/QCD~\cite{BrodskyPhysRep,
Trawinski}. Data points are from Ref.~\cite{pionff}.} 
\label{fig:pionff} 
\end{figure}

One represents the incoming pion as a bound state
of two quarks of constituent size $s_c$. According 
to Refs.~\cite{adsqft, Trawinski}, the effective 
constituent wave function should be a solution of 
a holographic eigenvalue equation~\cite{BrodskyPhysRep} 
with a quadratic effective potential, which for the 
ground state is Gaussian, $\psi_\pi \sim \exp{ 
[-\cM_{q \bar q}^2/(2\varkappa^2)] } $ where the 
constituent invariant mass squared is $\cM_{q \bar q}^2 
= (l^{\perp \, 2} + m_c^2)/[y(1-y)] = 4( \vec
l\,^2 + m^2_c)$. $\varkappa$ is a parameter of the 
quadratic effective potential written as $U_{\rm eff} 
= \varkappa^4 r^2/4$ in terms of the relative distance 
$r$ between the quark and antiquark, canonically conjugated to 
the relative momentum variables of Eqs.~(\ref{kperp})
and (\ref{kz}) according to quantum mechanics. The quark 
carries fraction $y$ and antiquark fraction $1-y$ of 
the pion large $+$-momentum. The pion transverse momentum 
is zero, and the quark carries transverse momentum 
$l^\perp$ while the antiquark carries $-l^\perp$. 
The wave function is normalized to 1, assuming that the 
effective quarks of size $s_c$ saturate the pion state, 
as it should be the case according to the classification 
of hadrons in particle data tables~\cite{pdg}. The spin 
and isospin details are described by multiplying the 
Gaussian factor by $\bar u ( p_\pi \hspace{-10pt}/\hspace{4pt}+ M) 
v$~\cite{thesis}. The wave function parameters are 
adjusted to reproduce the pion radius and decay constant
as well as the Gell-Mann--Okubo formula \cite{gell-mann, okubo}:
$m_c \sim 331$ MeV, $\varkappa \sim $ 436 MeV
and $M \sim - 1.9$ GeV. The pion radius squared is
$\sim 0.44$ fm$^2$ while data provide 0.45(1) 
fm$^2$~\cite{pdg}. Weak decay constant is $f_\pi \sim 
$ 130.7 MeV, to be compared with the experimental 
value of 130.4 MeV~\cite{pdg}. The fitted value of 
$\varkappa$ is about 25\% smaller than one could
expect on the basis of other models~\cite{Trawinski}.
The pion electromagnetic form factor obtained using 
these parameters is plotted in Fig.~\ref{fig:pionff}.

\begin{figure}[ht]
\centering
\includegraphics[scale=0.15]{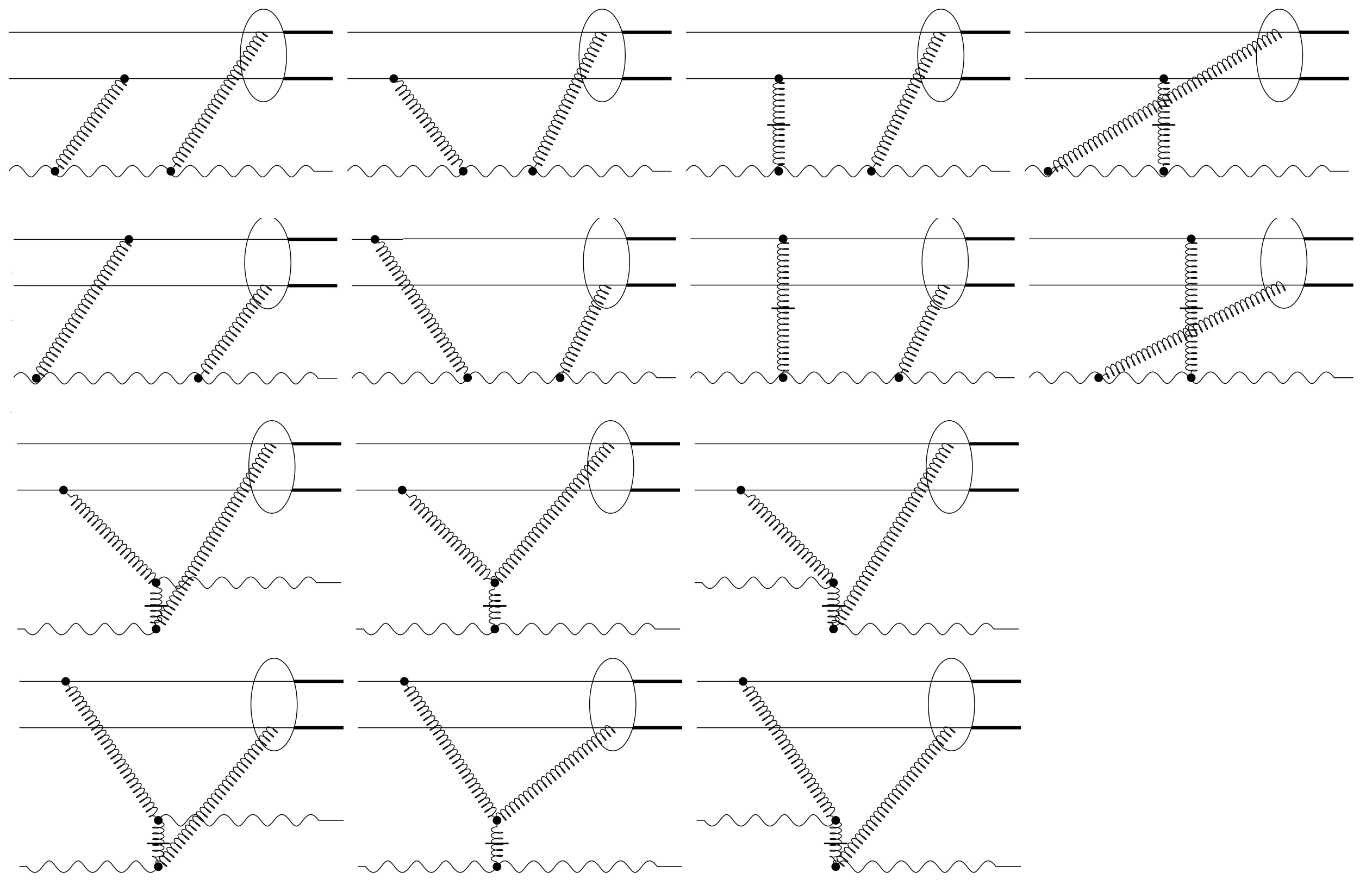}
\caption{ Examples of interactions that split the 
          pion on gluons in a nucleus according 
          to $H_{QCDs}$ with $1/s \sim 100$ GeV
          \cite{thesis}.}
\label{fig:piAgluons}
\end{figure}

\begin{figure}[ht]
\centering
\includegraphics[scale=0.5]{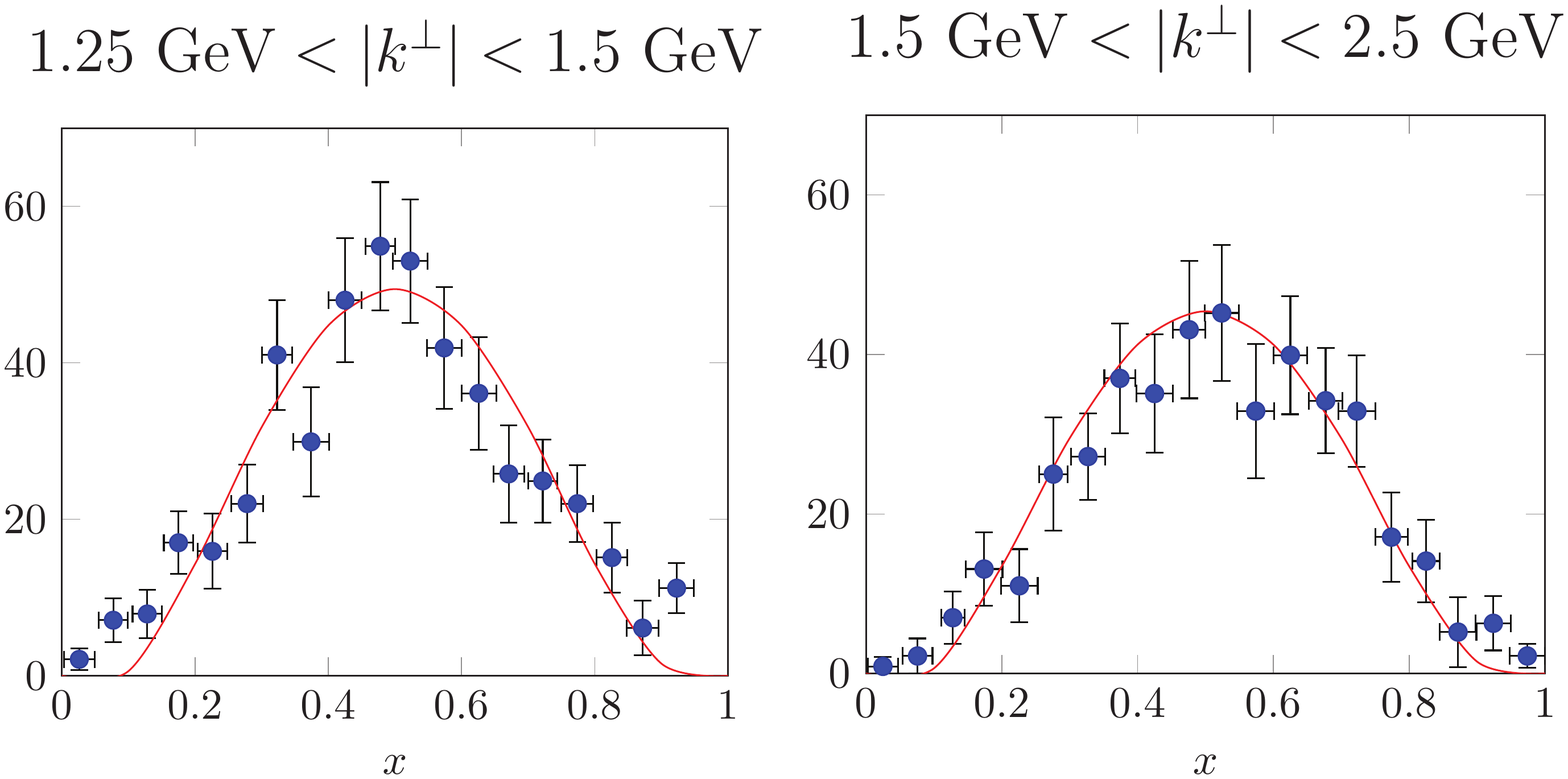}
\caption{Jet counts distribution $\phi(x,k^\perp)$, for jets 
         induced by pions impinging on Pt, in two jet-$k^\perp$ bins;
         data points are from Ref.~\cite{pionjets} and theory 
         curves from Ref.~\cite{thesis}.} 
\label{fig:jets}
\end{figure}

When the high-energy pion hits the nucleus and 
its constituents interact with the target, the
low-energy wave-function picture for the pion 
as built just from one constituent quark and 
one constituent 
\begin{figure}[ht]
\centering
\includegraphics[scale=0.3]{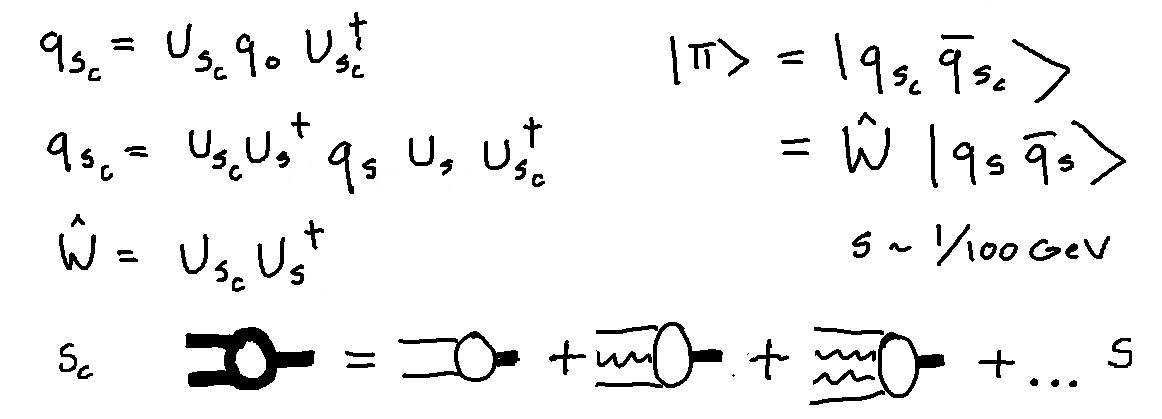}
\caption{ $|\pi\rangle = |q_{s_c} \bar q_{s_c}\rangle$ 
in terms of quarks and gluons with $1/s \sim 100$ GeV. }
\label{fig:qctoqs}
\end{figure}
antiquark is not adequate. We 
need to express the quarks of size $s_c$ by the 
quarks and gluons of much smaller size $s$ for 
which we know the nuclear distribution. For 
example, to use the platinum distribution of 
gluons in the Bjorken variable $x$ for $Q \sim
$ 100 GeV~\cite{gluons}, we need to use the RGPEP 
operator $U_s$ to transform the quarks from scale 
$s_c$ to the quarks of scale $ s \sim (100 \rm GeV)^{-1}$.
This is illustrated in Fig.~\ref{fig:qctoqs}. 
The required formula reads 
\beq
|\pi \rangle \es |q_{s_c} \bar q_{s_c} \rangle
             \rs \hat W_{s \, s_c} |q_s \bar q_s \rangle \ , 
\eeq
where the state $|q_s \bar q_s \rangle $ has the
same wave function as $|q_{s_c} \bar q_{s_c} \rangle $, 
$\hat W = U_{s_c} U^\dagger_s $ and both
operators $U$ are expressed in terms of creators and
annihilators of effective particles of size $s$.
The operator $\hat W$ can be evaluated using expansion
in powers of the coupling constant $g_s$, which is 
equivalent to expansion in powers of bare $g$ when
one limits the calculation to terms order 1, $g$ and 
$g^2$. To this order, the pion quark-antiquark 
constituent state has components with quark-antiquark, 
quark-antiquark-gluon and quark-antiquark-gluon-gluon 
of the small size $s$. Thus, the calculation of splitting 
of a pion into two jets amounts to evaluation of action
of $\hat W$ on the constituent model of pion suggested
by AdS/QCD-based holography and calculating scattering 
amplitude of the resulting components into final quarks, 
whose momenta are identified with the momenta of the 
outgoing jets. Examples of the diagrams that contribute 
are provided in Fig.~\ref{fig:piAgluons}.
The resulting jet counts distributions, $\phi(x,k^\perp)$ 
in Fig.~\ref{fig:pionA}, in comparison with data 
that are available in two bins of the jet $k^\perp$, 
are shown in Fig.~\ref{fig:jets}. The absolute 
normalization factor for the theoretical curves 
is freely adjusted in both bins, the required 
coefficients in low to high $k^\perp$ bin being of 
ratio 13/11. The shapes of theoretical curves 
are in a reasonably good agreement with experimentally 
observed distributions of jets. However, the calculation
does not include effects of propagation of leading quarks
through the nucleus and misses description of 
hadronization. Inclusion of these elements requires 
development of the RGPEP beyond the current stage.  

\section{ Conclusion }

The RGPEP opens new options for developing QFT
and applying it in phenomenology of particle,
nuclear and atomic physics. We are focused mainly 
on strong interactions. A critical step that 
awaits performing in QCD is to derive the Hamiltonian 
for effective particles with accuracy to fourth 
power of the coupling constant $g$. This Hamiltonian 
is expected to contain so far unknown terms whose
structure may shed new light on the form of effective 
dynamics that applies to hadrons understood in terms 
of constituent quarks. The simplest theory to start
with is QCD of quarks with masses much larger than 
$\Lambda_{QCD}$ in the RGPEP scheme. The goal would 
be to understand the dynamics of effective gluons 
in colorless objects, with the heavy quarks serving 
as anchors for the system. Extension to light
quarks will require understanding of how many of
them participate in the dynamics. This number will
be determined by the ratio of $\Lambda_{QCD}$ to 
the light quark masses, which is on the order of
100. Such large number explains why the answer is 
not simple to obtain. It involves understanding 
of the Hamiltonian mechanism of breaking chiral 
symmetry and buildup of confinement. The RGPEP
does not provide any ready answers to these
questions, but it does offer tools for required 
studies. In particular, the effective coupling 
constant $g_s$ at suitable values of quark and 
gluon size $s$ is not known yet and, so far, we do 
not have any theoretical estimates for the probability 
of finding large numbers of light quarks in eigenstates 
of $\hat H_{QCD \, s}$ with any value of $s$ that
is likely to be effective in describing light hadrons.

\begin{acknowledgements}
APT acknowledges support of the National Science 
Center of Poland grant PRELUDIUM under the decision 
number DEC-2014/15/N/ST2/03451.
\end{acknowledgements}

\end{document}